# Improving Scientific Workflow with Cloud Offloading

Hao Qian
hqianm@gmail.com

*Abstract* —Scientific workflow is a powerful tool to streamline and organize computational steps of scientific application. This paper presents Emerald, a system that adds sophisticated cloud offloading capabilities to scientific workflows. Emerald automatically offloads computation intensive steps of scientific workflow to the cloud in order to enhance workflow performance. Emerald provides easy-to-use APIs to help developers build cloud offloading enabled scientific workflows. Evaluation showed that Emerald can effectively reduce up to 55% of execution time for scientific applications.

*Keywords* — code offloading; scientific workflow; distributed computing; scheduling; cloud computing

## 1. Introduction

In scientific computing, workflow is used to organize and streamline steps of scientific computation. Formally speaking, scientific workflow is a specification of a scientific process, which represents, streamlines, and automates the analytical and computational steps that a scientist needs to go through from dataset selection and integration, computation and analysis, to final data product presentation and visualization [1].

Scientific workflow is a powerful paradigm for structuring and automating complex and distributed computation in various data-intensive sciences [2].

Scientific application is an increasingly critical foundation for research across the disciplines, from improving the foundations of linguistic analysis to protecting lives through better bullet-resistant vests. These applications are computation intensive, requiring high performance computing platforms like cluster or cloud platform. For organizations with limited budget, owning and maintaining a high performance computing infrastructure could be unrealistic. Cloud platform provides a scalable computing environment which helps reduce the cost of hosting scientific applications.

The computation demand of steps in a scientific workflow varies. Light weight computation steps can be executed by less powerful computer. In contrast, for computation steps with heavy computation, high performance computing platform like cloud are required. Many scientific workflow applications provide mechanisms to help developers build scientific workflows efficiently, but they don't provide mechanisms to determine where code is executed. We built Emerald to address this problem. Emerald provides services to offload computation intensive steps of a scientific workflow to the cloud. With Emerald, developers can build scientific workflows that are executed in a hybrid computing environment (i.e., local computer plus cloud platform). If a computation step should be offloaded to the cloud, developers only need to annotate it as *remotable.* At runtime, Emerald automatically offloads remotable steps to the cloud, greatly reducing the burden on developers to implement code offloading mechanism.

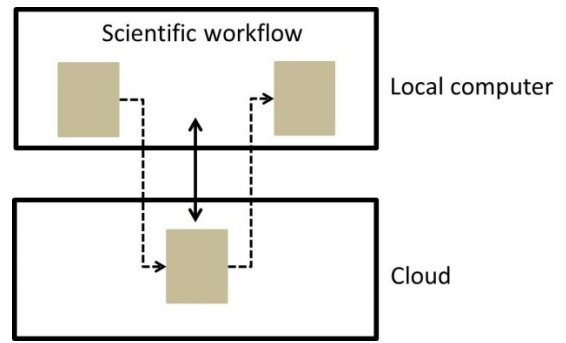

Figure 1: By offloading computation intensive steps of a workflow to the cloud, Emerald turns local execution of a workflow into distributed execution.

Our contributions are summarized as follows:
- We present the design and implementation of a complete system. Emerald is able to offload computation steps of scientific workflow from local computer to cloud platform.
- We present a Multi-level Data Storage Service (MDSS) that automatically and seamlessly synchronizes data between local computer and cloud. MDSS moves data to the cloud before computation offloading occurs in order to enhance the performance of scientific application.
- We evaluate Emerald with a real word scientific application. Results indicated that Emerald can effec-

tively enhance the performance of the application with seamless computation offloading.

## 2. System design

In this section, we present the high-level design of Emerald in order to demonstrate how they integrate into one system, thereby supporting distributed execution of scientific workflow.

Conceptually, Emerald automatically transforms scientific workflow execution on local computer into a distributed execution. Emerald is a flexible scientific workflow partitioner and execution runtime, it uses static analysis to partition scientific workflows developed by Windows Workflow Foundation (WF). At runtime, Emerald offloads computation intensive steps of a workflow from local computer to the cloud, executes it there, and re-integrates the migrated step back to the local computer (Figure 1).

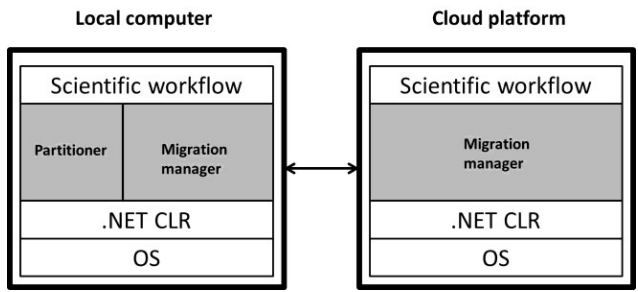

Figure 2: Components of Emerald

In order to increase understanding of this system, terms used in Emerald must be defined. A scientific application contains *computation steps*. Developers can annotate a computation step as *remotable step*, meaning the computation step can be offloaded to the cloud. If a remotable step is executed on local computer (i.e., it is not offloaded), we call it *local execution*. In contrast, if a remotable step is executed on the cloud (i.e., it is offloaded), we call it *remote execution*. In this paper, we assume local computer are resource constraint computers, whereas the cloud are resource rich, powerful servers. Emerald components are shown in Figure 2.

The implementation details of Emerald are provided in section 3.

## 3. Implementation

### 3.1 Workflow partition

In Windows Workflow Foundation (WF), workflow is defined by XAML (Extensible Application Markup Language) file. Each step of workflow is represented by a node with corresponding properties. The hierarchical structure of XAML file makes it easy to analyze the relationship of steps. Figure 3 shows a sample WF workflow.

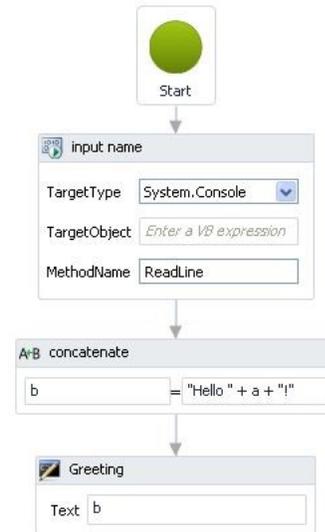

Figure 3: Sample WF workflow

The workflow consists of three steps: input name, concatenate and greeting. It asks user to input name, then concatenates "Hello" with user's name to form a greeting message, finally shows the message on the screen. The XAML file representing this workflow is:

```
<Flowchart.StartNode>

    <InvokeMethod DisplayName="input name">
    </InvokeMethod>

    <Assign DisplayName="concatenate">
    </Assign>

    <WriteLine DisplayName="Greeting">
    </WriteLine>

</Flowchart.StartNode>
```

The root node of the XAML file represents the workflow, like XML, a XAML document must have one root node. The three child nodes represent three steps in the workflow. Each node can also have child nodes which forms a nested workflow.

The choice of which computation step to offload is made by workflow developer by adding one attribute to the corresponding XAML code. If a step is suitable for offloading, developer only needs to add a migration attribute to the node (Figure 4). Computation step with migration attribute will be recognized, partitioned and offloaded to the cloud at runtime by Emerald.

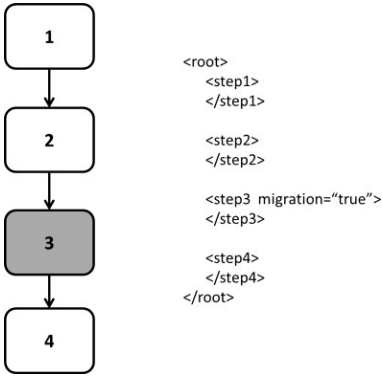

Figure 4: Emerald offloads computation step with migration attribute to the cloud.

The Emerald partitioner aims to determine which part of the workflow to retain on the local computer and which part to offload to the cloud. Any WF scientific workflow annotated according to the rules can be partitioned. The partitioner reads the annotated workflow and outputs a modified workflow with migration points (Figure 5).

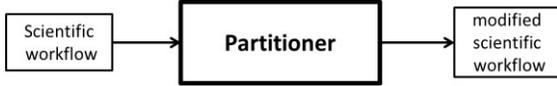

Figure 5: Given an annotated workflow as input, the Emerald partitioner outputs a modified workflow with migration points.

A partitioning example is shown in Figure 6. When partitioner analyzes the annotated XAML file, remotable step B is found, a temporary step is inserted before it. The job of the temporary step is to suspend the execution of the workflow and notify migration manager to offload step B. After step B returns from the cloud, the temporary step resumes the execution of the workflow.

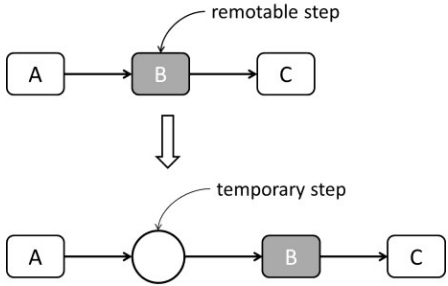

Figure 6: A temporary step is inserted before the remotable step

### 3.2 Constraints

To partition a workflow, developers need to follow some rules. Three properties of any legal partition are explained here:

*Property 1. Steps that access special hardware of the local computer can't be offloaded.*

If a step uses special resources such as GPU or other hardware accelerator, the step must remain locally. This property guarantees that the partitioned workflow has a better compatibility with cloud platform.

*Property 2. The input and output data of a step must be defined as variables of the workflow, and should be in the same level with the step.*

In WF, step can contain variables, and variable has scope. If a variable is defined in a step, it is available to the step and its nested workflows. In the workflow shown in Figure 7, there are three variables: A, B and C. A is defined in step 1, it is accessible from step 1 and nested step a and b. B is defined in step a, it is only accessible from step a, sibling step b and parent step 1 can't access B. C is defined in the same scope as step 1 and step 2, so all steps in the workflow can access C.

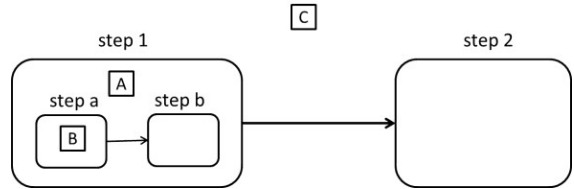

Figure 7: Variables have different scope in WF

In a scientific workflow, the output of the upstream step may become input for the downstream step. To make sure data can be shared between different steps, the input and output data should be defined in the same level as steps (Figure 8).

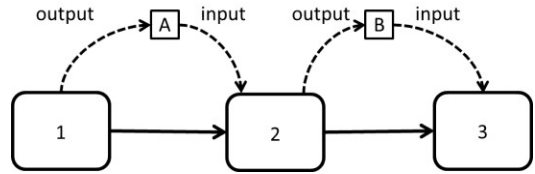

Figure 8: Input and output data should be defined in the same level as steps, so data is accessible to all steps.

*Property 3. Nested offloading is not allowed.*

This implies that nested suspends and nested resumes are not allowed. Once the workflow is suspended for offloading at some point, the workflow should not suspend again before resume, i.e., migration and re-integration should happen alternately.

### 3.3 Distributed Execution

The goal of the distributed execution mechanism of Emerald is to offload a step of workflow as determined by the partitioner from local computer to the cloud, execute it there, and re-integrate it back to local computer. The migration manager handles the distributed execution.

The life-cycle of the distributed execution is as follows. After the modified workflow is created by the partitioner, the user can launch the workflow on local computer. When the execution of the workflow reaches a remotable step $i$, the temporary step before $i$ suspends the execution of the workflow, notifies the migration manager that $i$ should be offloaded to the cloud. The migration manager records the information of $i$ and offloads it to the cloud. The cloud's migration manager receives $i$ and resumes its execution. When $i$ finishes execution, it is packaged as before and shipped back to the local computer. Finally, $i$ is merged into the original workflow.

Emerald operates at the granularity of computation step, allowing parallel steps to be offloaded and executed concurrently on the cloud. In a sequential workflow, the downstream step needs to wait until upstream step finishes to start execution. In a parallel workflow, parallel steps don't affect each other, they can be executed concurrently (Figure 9).

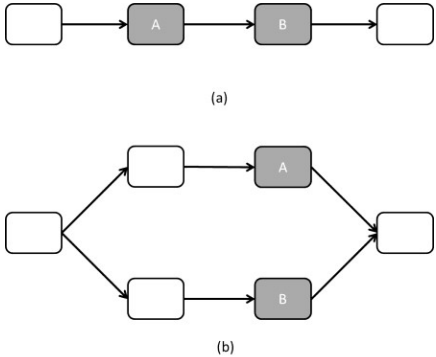

Figure 9: (a) Computation steps are executed in sequence in sequential workflow. (b) Parallel steps can be offloaded and executed concurrently.

*3.4 Multi-level data storage*

Emerald provides a Multi-level Data Storage Service (MDSS) that optimizes the performance of scientific workflow when computation offloading occurs. MDSS allows developers to save application data on the cloud without writing any backend code. Application data is also saved locally allowing applications to work offline. MDSS automatically synchronizes data between local computer and the cloud, so developers can focus on creating applications instead of having to worry about building backend solution to handle data storage and synchronization.

When application generates new data, MDSS first saves the data on local computer, so data is always accessible to application. Data is uploaded to the cloud later when MDSS performs synchronization. Synchronization of data sets between local computer and the cloud can be triggered by calling the *synchronize* method. In order to synchronize, MDSS reads the latest version of the data available in the cloud and compares it to the local copy. After comparison, MDSS writes the latest updates as necessary to the local copy and the cloud. MDSS maintains the last-written version of the data by default.

A remotable step usually contains two elements: 1) application data (e.g., images, texts, numbers) and 2) task code that performs execution on application data (e.g., sorting algorithm, image processing). Application data and task code are bundled and transferred when a remotable step is offloaded to the cloud. In most cases, the size of application data is much bigger than the size of task code (e.g., size of an image could be a few MB, whereas size of task code performing complex computation could be a few KB). By introducing MDSS, a remotable step's application data and task code are separated. In Emerald, a remotable step $i$ contains only task code, the application data accessed by $i$ is stored separately and referenced by URI. When $i$ is offloaded to the cloud, if the cloud already has the most recent copy of the application data that $i$ needs to access, Emerald only offloads task code to the cloud in order to reduce the amount of data transferred. MDSS effectively helps reducing the amount of data in the network by avoiding transferring application data every time when $i$ is offloaded.

Emerald uses URI to reference the application data to be acted on. When a remotable step $i$ is chosen for offloading, URI of $i$'s data is passed to MDSS which then queries the data using the URI. If the latest version of the application data is found on the cloud, Emerald offloads $i$ to the cloud without synchronization. If the cloud does not have the application data or the cloud has an older version of the application data, MDSS synchronizes the cloud with local computer before offloading $i$ (Figure 10).

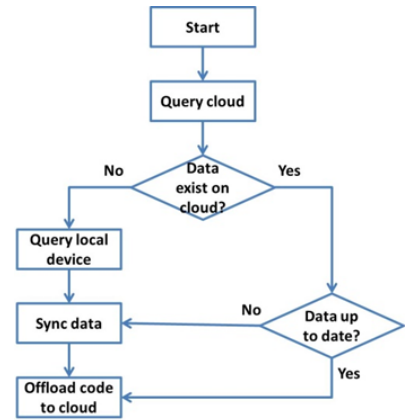

Figure 10: MDSS reduces the amount of data transferred in the network if the cloud already has the latest application data.

## 4. Evaluation

In this section we evaluate Emerald's ability to enhance the performance of scientific workflow by offloading computation intensive steps to the cloud.

A scientific application for Adjoint Tomography (AT) was used for the evaluation. The goal of adjoint tomography is to get higher resolution image of the earth structure through solving 3D seismic wave equation. The higher resolution earth structure image can provide additional constraints for geological interpretation. There are many applications of adjoint tomography, for example, looking for geothermal energy. Geothermal energy is clean, renewable and sustainable resource. Using geothermal energy can reduce the greenhouse gas emission resulting from burning fossil fuel. Another application of adjoint tomography is to help image the detail of underground earth structure to find safe structure for $CO_2$ sequestration.

AT includes four computational steps: (1) it builds a starting model and calculates synthetic seismograms based on the model; (2) it compares the synthetic seismograms with the observed data to find the misfit measurements; (3) it generates model perturbation by calculating Frechet Kernel; and (4) it generates an update model by applying model perturbation to the starting model. These steps are repeated until the seismograms generated by the update model can match the observed data wiggle by wiggle. AT can generate accurate and high resolution image by solving the 3D wave equation accurately compared with the classic ray theory which approximates the wave in a line. In our evaluation, step 2, 3 and 4 were annotated as remotable step in order to be offloaded to the cloud.

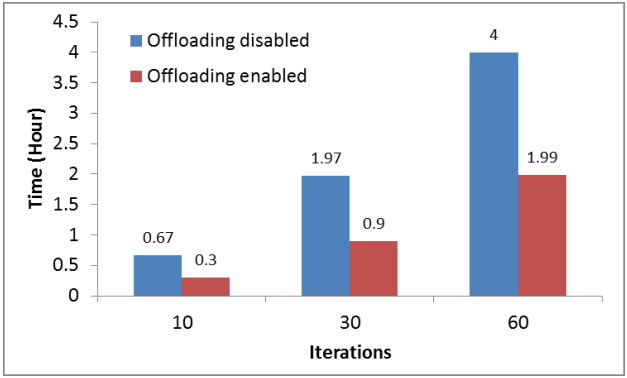

Figure 11: Execution time of AT on 104×23×24 mesh

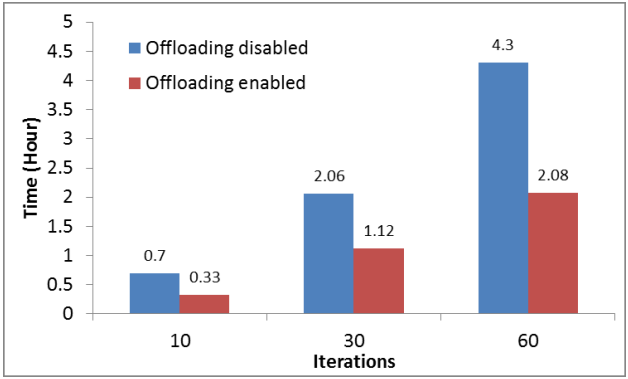

Figure 12: Execution time of AT on 208×44×46 mesh

An Emerald scientific workflow was developed to automate the execution of AT. The workflow was deployed on a local cluster. Ten computational nodes were used to run AT. The hardware configuration of each node is:
- One Intel Xeon quad core processors running at 3.2GHz, 48GB RAM
- Three nodes have seven Nvidia Fermi C2050 GPU cards.
- The file server has approximately 28TB of shared disk space.

For computation offloading, we created 25 D-series virtual machines on Microsoft Azure cloud platform. Each virtual machine has 16 cores, 112GB RAM and 800GB SSD.

In order to evaluate how cloud platform can enhance AT's performance, for each input (mesh), we ran AT twice: first time with computation offloading disabled, AT was executed on local cluster; and second time with computation offloading enabled, for each iteration, step 2, 3 and 4 were offloaded to the cloud. Before the experiment, AT's data were synchronized between local cluster and the cloud in order to reduce data transfer overhead.

The results showed that Emerald can help scientists create scientific workflow efficiently, and during workflow's execution, Emerald automatically moves remotable steps to the cloud seamlessly. When computation offloading was enabled, the execution time of AT can be reduced up to 55% (Figure 11 and 12).

## 5. Related Work

A scientific computational experiment often spans multiple computational and analytical steps, and during the execution, researchers need to store, access, transfer, and query information. Scientific workflow is a powerful tool to streamline and organize scientific application. Numbers of tools have been developed to help scientists build scientific workflow.

There have been many efforts to utilize scientific workflow for domain-specific problems. In [3], a scientific workflow was developed to automate the inside-out process of enzyme design. It was developed using Kepler and deployed on the grid. By distributing the time-consuming parts of the application to computational grid, the workflow system speeded up the enzyme design process and provided ability to store, access, transfer, and query information during the execution.

In [1], Great Lake Forecasting System (GLFS) is a scientific application that monitors meteorological conditions of the lake Erie for nowcasting (for next hour) and forecasting (for next day). A workflow system was used to run GLFS, and it was deployed on clusters. Every second, the input data was coming from satellites supervising this area and sensors planted along the costal line.

These are good examples of using scientific workflow to automate and speed up scientific application. Problems of those solutions include:
- The workflow is developed by specific tool, such as Kepler, so it is not compatible with components developed by other tools. If we need to add some existing functionality developed by other tools into the workflow, many changes need to be done.
- As cloud computing is emerging as an attracting platform, it provides better scalability and flexibility to deploy scientific applications, and it reduces the infrastructure investment for organizations. It is hard to move such existing scientific workflows onto the cloud, because they are developed for either cluster or grid, and many scientific development tools don't support cloud computing.

Emerald addresses these problems. Based on WF, workflow developed by Emerald can integrate with any components developed by other tools. Emerald provides services for offloading computation to the cloud, thereby, developers can easily add cloud computing ability to workflows with little effort.

Emerald was also built upon previous research regarding program partitioning, code offloading, and remote execution [4] and [5]. Computation offloading is an effective method to alleviate the restrictions of local computer by sending heavy computations to resourceful servers and receiving results from these servers [6] and [7]. Many issues related to computation offloading have been investigated in the past decade, including feasibility of offloading, offloading decisions, and development of offloading infrastructures.

## 6. Future Work

In the future works, we will study more sophisticated data placement strategies between cloud and local computer to further reduce the data transfer overhead.

Another direction of future work is security [12]. Security concerns arise when code is offloaded to servers, for example, multiple applications accessing a single server, running foreign code on the server, and remote codes interfering with each other. We will study how to protect the security of applications when code offloading occurs.

## 7. Conclusion

In this paper, we presented Emerald, a scientific workflow application that helps developers create scientific workflows efficiently and enables computation offloading to the cloud. Emerald can effectively enhance the performance of scientific workflow by offloading computation intensive steps to powerful cloud platform.

We evaluated Emerald with a real world adjoint tomography scientific application. Results showed that Emerald can effectively reduce the application's execution time by up to 55%.